\documentstyle[12pt,a4]{article}

\newcommand{\mb}[1]    {\ifmmode#1\else\mbox{$#1$}\fi}
\newcommand{\as}       {\mb{\alpha_s}}
\newcommand{\calA}     {\mb{{\cal A}}}
\newcommand{\calO}     {\mb{{\cal O}}}
\newcommand{\calM}     {\mb{{\cal M}}}
\newcommand{\ee}       {\mb{\mathrm{e^+e^-}}}
\newcommand{\beq}      {\begin{equation}}
\newcommand{\eeq}      {\end{equation}}
\newcommand{\beqn}     {\begin{eqnarray}}
\newcommand{\eeqn}     {\end{eqnarray}}
\newcommand{\slsh}     {\rlap{$\;\!\!\not$}}
\newcommand{\ldot}[2]  {\:{#1}\!\cdot\!{#2}\:}
\newcommand{\myfrac}[2]{\mbox{\small$\frac{#1}{#2}$}}
\newcommand{\half}     {\myfrac{1}{2}}
\newcommand{\sss}[1]   {{\scriptscriptstyle\mathrm{#1}}}
\newcommand{\sssB}     {{\sss B}}
\newcommand{\sssF}     {{\sss F}}
\newcommand{\VEV}[1]   {\mb{\left\langle #1\right\rangle}}

\catcode`\@=11
\newlength{\captionsize}
\newlength{\captionlength}
\long\def\@caption#1[#2]#3{\par\addcontentsline{\csname
  ext@#1\endcsname}{#1}{\protect\numberline{\csname
  the#1\endcsname}{\ignorespaces #2}}\begingroup
    \@parboxrestore
    \normalsize
    \@makecaption{\csname fnum@#1\endcsname}{\ignorespaces
\settowidth{\captionsize}{#1~\csname the#1\endcsname:}%
\setlength{\captionsize}{-\captionsize}%
\addtolength{\captionsize}{\textwidth}%
\addtolength{\captionsize}{-3mm}%
\settowidth{\captionlength}{#3}%
\ifdim\captionlength<\captionsize{#3}\else%
\parbox[t]{\captionsize}{#3}%
\fi%
}\par
  \endgroup}
\catcode`\@=12

\begin{document}
\begin{titlepage}
\addtocounter{page}{1}
\begin{flushright}
     LU TP 94-13 \\
     hep-ph/9410244 \\
     September 1994
\end{flushright}
\par \vskip 15mm
\vspace*{\fill}
\begin{center}
\Large\bf
                A Simple Prescription for First Order \\
                 Corrections to Quark Scattering and \\
                       Annihilation Processes%
\footnote{Work supported in part by the EEC Programme ``Human Capital
and Mobility'', Network ``Physics at High Energy Colliders'', contract
CHRX-CT93-0537 (DG 12 COMA).}
\end{center}
\par \vskip 2mm
\begin{center}
        {\bf Michael H.\ Seymour}%
\footnote{Address after 1st January 1995: CERN TH Division, CH-1211
Geneva 23, Switzerland.} \\
        Department of Theoretical Physics, University of Lund, \\
        S\"olvegatan 14A, S-22362 Lund, Sweden
\end{center}
\par \vskip 2mm

\begin{center} {\large \bf Abstract} \end{center}
\begin{quote}
\pretolerance 1000 
We formulate the first order corrections to processes involving the
scattering or annihilation of quarks in a form in which the QCD and
electroweak parts are exactly factorised.  This allows for a
straightforward physical interpretation of effects such as lepton-hadron
correlations, and a simpler Monte Carlo treatment.
\end{quote}
\vspace*{\fill}
\begin{flushleft}
     LU TP 94-13 \\
     September 1994
\end{flushleft}
\end{titlepage}
\section{Introduction}

Although the parton model has been very successful in providing a
qualitative description of \ee\ annihilation to hadrons, deep inelastic
lepton-hadron scattering (DIS), and Drell-Yan lepton pair production, a
more detailed description requires higher-order QCD corrections.  These
have been known for many years to first order in \as, and more recently
to second order, but are generally presented in an inclusive form.
However, in many instances, a more exclusive treatment is needed, for
example to describe features of the hadronic final state or correlations
between the hadronic and leptonic parts of the process.  This is
particularly true of Monte Carlo event generators, where the aim is to
provide a fully exclusive description of the process, event by event.

In this paper we present the first order tree-level corrections to quark
scattering and annihilation processes in a form in which the QCD and
electroweak parts {\em exactly} factorise.  This makes exclusive event
features and correlations particularly transparent.  We also discuss how
these could be used to provide a simple Monte Carlo treatment of the
processes.  We essentially follow the method of Ref.~[\ref{K}] for \ee\
annihilation, taking over most of the same notation.

In anticipation of the main applications of our results, we use the
language of the DIS and Drell-Yan processes, but in fact our method is
applicable to any electroweak process in which the lowest-order diagrams
contain a single quark line attached to a single gauge boson.  It should
also be stressed that the method treats the quarks as massless
throughout.

The paper is set out as follows: In the remainder of this section, we
recap the important ingredients of [\ref{K}].  In section~\ref{QCDC} we
follow the same method to derive the equivalent expression for the QCD
Compton part of the first-order correction to DIS (QCDC,
$\mathrm{q\ell\to qg\ell}$).  In section~\ref{BGF} we do the same for
the boson gluon fusion part (BGF, $\mathrm{g\ell\to q\bar{q}\ell}$).  In
section~\ref{azi} as an example of the simplicity of our form, we
discuss lepton-hadron correlations in DIS, which have been proposed as
an important test of QCD[\ref{GP},\ref{CES}].  In section~\ref{DY} we
derive and discuss the equivalent results for the Drell-Yan process.
Finally in section~\ref{summ} we give a summary.

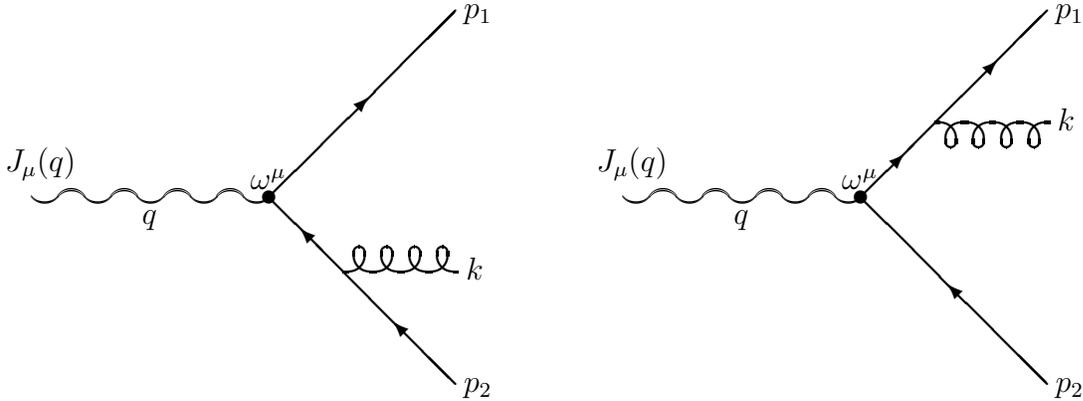
\begin{figure}
\input FEYNMAN
\centerline{
\begin{picture}(22000,18000)
\pbackx=11000\pbacky=9000
\THICKLINES
\bigphotons
\drawline\photon[\W\REG](\pbackx,\pbacky)[9]
\global\advance\pmidx by -300
\global\advance\pmidy by -1000
\put(\pmidx,\pmidy){$q$}
\global\advance\pbackx by -1000
\global\advance\pbacky by 1000
\put(\pbackx,\pbacky){$J_\mu(q)$}
\drawline\fermion[\NE\REG](\pfrontx,\pfronty)[10000]
\drawarrow[\NE\ATBASE](\pmidx,\pmidy)
\global\advance\pbackx by  300
\global\advance\pbacky by -300
\put(\pbackx,\pbacky){$p_1$}
\drawline\fermion[\SE\REG](\pfrontx,\pfronty)[4000]
\drawarrow[\NW\ATBASE](\pmidx,\pmidy)
\put(\pfrontx,\pfronty){\circle*{500}}
\global\advance\pfrontx by -700
\global\advance\pfronty by 300
\put(\pfrontx,\pfronty){$\omega^\mu$}
\drawline\gluon[\E\FLIPPED](\pbackx,\pbacky)[4]
\global\advance\pbackx by 300
\global\advance\pbacky by -300
\put(\pbackx,\pbacky){$k$}
\drawline\fermion[\SE\REG](\pfrontx,\pfronty)[6000]
\drawarrow[\NW\ATBASE](\pmidx,\pmidy)
\global\advance\pbackx by 300
\global\advance\pbacky by -300
\put(\pbackx,\pbacky){$p_2$}
\end{picture}
\hfill
\begin{picture}(22000,18000)
\pbackx=11000\pbacky=9000
\THICKLINES
\bigphotons
\drawline\photon[\W\REG](\pbackx,\pbacky)[9]
\global\advance\pmidx by -300
\global\advance\pmidy by -1000
\put(\pmidx,\pmidy){$q$}
\global\advance\pbackx by -1000
\global\advance\pbacky by 1000
\put(\pbackx,\pbacky){$J_\mu(q)$}
\drawline\fermion[\SE\REG](\pfrontx,\pfronty)[10000]
\drawarrow[\NW\ATBASE](\pmidx,\pmidy)
\global\advance\pbackx by  300
\global\advance\pbacky by -300
\put(\pbackx,\pbacky){$p_2$}
\drawline\fermion[\NE\REG](\pfrontx,\pfronty)[4000]
\drawarrow[\NE\ATBASE](\pmidx,\pmidy)
\put(\pfrontx,\pfronty){\circle*{500}}
\global\advance\pfrontx by -700
\global\advance\pfronty by 300
\put(\pfrontx,\pfronty){$\omega^\mu$}
\drawline\gluon[\E\REG](\pbackx,\pbacky)[4]
\global\advance\pbackx by 300
\global\advance\pbacky by -300
\put(\pbackx,\pbacky){$k$}
\drawline\fermion[\NE\REG](\pfrontx,\pfronty)[6000]
\drawarrow[\NE\ATBASE](\pmidx,\pmidy)
\global\advance\pbackx by 300
\global\advance\pbacky by -300
\put(\pbackx,\pbacky){$p_1$}
\end{picture}
}
\label{e+e-}
\caption{Tree-level Feynman diagrams for $\ee\to$ hadrons at
$\calO(\as),$ in which the lepton side is represented by an arbitrary
current, $J_\mu,$ and the boson-quark coupling by $\omega^\mu$.}
\end{figure}

The tree-level Feynman diagrams for \ee\ annihilation to hadrons at
$\calO(\as)$ are shown in Fig.~\ref{e+e-}.
It is well known that only the sum of the two diagrams is gauge
invariant, and that they can never be separated.  However, the first
important step of [\ref{K}] is to use the explicit gauge choice
introduced by the CALKUL collaboration[\ref{CALKUL}], in which the
diagrams only contribute to physically distinct processes, with all the
interference absorbed into the form of the polarisation tensor.  In this
gauge, only the first diagram of Fig.~\ref{e+e-} contributes to final
states in which the gluon has the same helicity as the quark, and the
second diagram when it has the same helicity as the antiquark.  The
polarisation tensor contains collinear-divergent pieces, so that both
diagrams still give collinear divergences in both directions.

The next important step is to introduce a pair of massless four-vectors
$r_{1,2}$ that are parallel to $p_{1,2}$ but with energy $Q/2,$ the
energy of the partons in the parton model.  These allow each matrix
element to be written as the product of two parts, one corresponding to
parton-model production, the other to the emission of the gluon without
any knowledge of $J_\mu$ or $\omega^\mu$.  The final result is then
\beq
  d\sigma_3 = \frac{C_\sssF\as}{2\pi} d\Gamma_2
              \frac{dx_1dx_2}{(1-x_1)(1-x_2)} \frac{d\phi}{2\pi}
  \left\{x_1^2|\calM_2(r_1,q-r_1)|^2+
         x_2^2|\calM_2(q-r_2,r_2)|^2\right\}\!,
\label{ee}
\eeq
where $x_i\equiv2\ldot{p_i}{q}/\ldot{q}{q}$ are the energy fractions in
the cmf of $q,\;d\Gamma_2$ is an element of the parton-model
phase-space, and $\phi$ is the azimuthal angle of the gluon.  Both
$d\Gamma_2$ and $\phi$ have different interpretations for the two parts
of the matrix element: in the first they refer to the phase-space to
produce a quark with momentum $r_1$ in the parton model, and the
azimuthal angle around $r_1,$ while in the second they refer to an
antiquark with momentum $r_2$.  With the exception of neglecting quark
masses, this expression is exact for arbitrary currents of vector or
axial-vector type, so applies equally well to the W decays in WW pair
production for example.  It is also valid at the helicity level, so if
polarisation effects are retained in the lowest-order matrix element,
they are correctly treated at $\calO(\as)$.

As a Monte Carlo prescription, (\ref{ee}) has a simple interpretation:
first parton-model events are generated according to the exact matrix
element, then gluon emission is generated according to
\beq
  \frac1{\sigma_0}\frac{d\sigma}{dx_1dx_2} = \frac{C_\sssF\as}{2\pi}
              \frac{x_1^2+x_2^2}{(1-x_1)(1-x_2)},
\eeq
and finally parton 1 or 2 is chosen with relative probability $x_1^2$
and $x_2^2$ to retain its parton-model direction, with the hadron plane
being rotated uniformly in azimuth about it.  The quarks retain their
lowest-order polarisation states, with the gluon inheriting the same
state as the parton that was chosen.

This procedure produces the exact distribution of three-parton final
states, with no knowledge of the mechanism that produced them, provided
an exact generator of two-parton final states is available.

\section{QCD Compton Process in DIS}
\label{QCDC}

\subsection{The Matrix Element}

The three-parton matrix element for a quark of helicity $\lambda$ and a
gluon of helicity $\rho$ is given by
\beq
  \calM_3(\lambda,\rho) = ig_s\bar{u}_\lambda(p_2)
  \left[\slsh\epsilon_\rho(k)\frac{-(\slsh p_2+\slsh k)}{2\ldot{p_2}{k}}
        \omega^\mu+\omega^\mu\frac{-(\slsh p_1-\slsh k)}{-2\ldot{p_1}{k}}
        \slsh\epsilon_\rho(k)\right]u_\lambda(p_1)J_\mu(q),
\eeq
where the momentum of the incoming(outgoing) quark is $p_1(p_2)$, and
the gluon is $k$.  All the partons are considered massless.  We use the
explicit formulation of the gluon polarisation tensor $\epsilon_\rho(k)$
introduced by the CALKUL collaboration,
\beqn
  \slsh\epsilon_\rho(k) &=& N\left(
    \half(1+\rho\gamma^5)\slsh k\slsh p_2\slsh p_1 -
    \slsh p_2\slsh p_1\slsh k\half(1+\rho\gamma^5) \right), \\
  N &=& \left[ 4 \ldot{p_1}{k} \ldot{p_2}{k} \ldot{p_1}{p_2}
    \right]^{-\half}.
\eeqn
After a little manipulation, we obtain
\beqn
  \calM_3^+\equiv\calM_3(\lambda,\phantom{-}\lambda) &=&
    -ig_sN\bar{u}_\lambda(p_2)\omega^\mu(\slsh p_1-\slsh k)\slsh p_2
    u_\lambda(p_1)J_\mu(q), \\
  \calM_3^-\equiv\calM_3(\lambda,-\lambda) &=&
    -ig_sN\bar{u}_\lambda(p_2)\slsh p_1(\slsh p_2+\slsh k)\omega^\mu
    u_\lambda(p_1)J_\mu(q).
\eeqn
The special vectors we introduce are defined covariantly by the momentum
fractions of the partons,
\beq
  x_i \equiv \frac{2\ldot{p_i}{q}}{\ldot{q}{q}},
\eeq
just as in \ee\ annihilation.  In general $x_1\le-1,\;x_2\le1,$ where
the equalities apply in the parton-model scattering case.
\beqn
  r_1       &\equiv&          - p_1/x_1,\\
  \bar{r}_2 &\equiv& \phantom{-}r_1+q=p_2+k-p_1-p_1/x_1,\\
  r_2       &\equiv& \phantom{-}p_2/x_2,\\
  \bar{r}_1 &\equiv& \phantom{-}r_2-q=p_1-k-p_2+p_2/x_2.
\eeqn
It can be seen that $r_1,\bar{r}_2$ are the momenta that the incoming
and outgoing quarks would have in $\calO(1)$ scattering at the same
$y_\sssB$ and $Q^2$. $\bar{r}_1,r_2$ have the same component parallel to
the current direction, but also have a component transverse to it, such
that $r_2$ is parallel to $p_2$.  As we discuss in more detail later,
these are the momenta that the incoming and outgoing quarks would have
in the parton model with an `intrinsic' $p_t$.

We can use these vectors to rewrite the matrix elements using the
following results,
\beqn
  u(\alpha p) &\cong& \sqrt\alpha u(p), \\
  (\slsh p_1-\slsh k)\slsh p_2 &=& \slsh\bar{r}_1\slsh p_2 \\
  &=& u_\lambda(\bar{r}_1)\bar{u}_\lambda(\bar{r}_1)\slsh
    p_2, \\
  \slsh p_1(\slsh p_2+\slsh k) &=&
    \slsh p_1u_\lambda(\bar{r}_2)\bar{u}_\lambda(\bar{r}_2),
\eeqn
where $\cong$ denotes ``equal except for an overall complex phase'', and
all the results rely on the assumed masslessness of $p_{1,2}$.
Inserting these into the matrix elements we obtain
\beqn
  \calM_3^+ &\cong& g_sN\sqrt{x_2}\;\bar{u}_\lambda(r_2)
    \omega^\mu u_\lambda(\bar{r}_1)\bar{u}_\lambda(\bar{r}_1)\slsh p_2
    u_\lambda(p_1)J_\mu(q) \\
  &\equiv& C^+\calM_2(\bar{r}_1,r_2) ,\\
  C^+ &=& g_sN\sqrt{x_2}\;\bar{u}_\lambda(\bar{r}_1)
    \slsh p_2u_\lambda(p_1), \\
  \calM_3^- &\equiv& C^-\calM_2(r_1,\bar{r}_2) ,\\
  C^- &=& g_sN\sqrt{x_1}\;\bar{u}_\lambda(p_2)
    \slsh p_1u_\lambda(\bar{r}_2),
\eeqn
where $\calM_2(q_1,q_2)=\bar{u}_\lambda(q_2) \omega^\mu u_\lambda(q_1)
J_\mu(q)$ is the parton-model matrix element for the current to scatter
an incoming quark $q_1,$ to an outgoing quark $q_2$.  Finally we can
calculate the $|C^{\pm}|^2$ explicitly to give
\beqn
  |C^+|^2 &=& \frac{8\pi\as}{(-1-x_1)(1-x_2)Q^2} x_2^2 ,\\
  |C^-|^2 &=& \frac{8\pi\as}{(-1-x_1)(1-x_2)Q^2} x_1^2 .
\eeqn

\subsection{Kinematics and Phase-Space}

We parametrise the parton model scattering by $x_\sssB,\;Q^2$ and
$\Phi,$ the lab-frame azimuth of the lepton scattering, which is uniform
for leptons with no transverse polarisation.  In addition, we
parametrise the hadron-plane momenta by $x_{1,2}$~and~$\phi,$ the
azimuth of the outgoing quark around the boson-hadron axis in their cmf
(or, equivalently, the Breit frame).  To be precise, when discussing the
specific case of DIS, $\phi=0$ when the outgoing lepton and quark are
parallel.  The phase-space limits are
\beqn
  \frac{-1}{x_\sssB} \;<& x_1 &<\; -1, \\
  1+x_1              \;<& x_2 &<\; \phantom{-}1.
\eeqn
The cross-section at $\calO(1)$ is then
\beqn
  d\sigma_2 &=& \frac{1}{64\pi}\;\frac{1}{s^2x_\sssB^2}f(x_\sssB,Q^2)
    dQ^2dx_\sssB\frac{d\Phi}{2\pi}\;\Sigma|\calM_2|^2 \\
  &\equiv& d\Gamma_2\Sigma|\calM_2|^2,
\eeqn
and at $\calO(\as)$,
\beqn
  d\sigma_3 &=& \frac{C_\sssF}{128(2\pi)^3}\;\frac{1}{s^2x_\sssB^2}
    f(-x_\sssB x_1,Q^2)
    dQ^2dx_\sssB\frac{d\Phi}{2\pi}\frac{dx_1}{x_1^2}dx_2\frac{d\phi}{2\pi}\;
    Q^2\Sigma|\calM_3|^2 \\
  &=& \frac{C_\sssF}{4(2\pi)^2}d\Gamma_2
    \frac{-x_\sssB x_1f(-x_\sssB x_1,Q^2)}
         { x_\sssB    f( x_\sssB    ,Q^2)}
    \frac{dx_1}{-x_1^3}dx_2
    \frac{d\phi}{2\pi}\;
    Q^2\Sigma|\calM_3|^2 \\
  &=& \frac{C_\sssF\as}{2\pi}d\Gamma_2
    \frac{-x_\sssB x_1f(-x_\sssB x_1,Q^2)}
         { x_\sssB    f( x_\sssB    ,Q^2)}
    \frac{dx_1dx_2}{-x_1^3(-1-x_1)(1-x_2)}\;\frac{d\phi}{2\pi}
\nonumber\\&&
   \left\{ x_1^2|\calM_2(r_1,\bar{r}_2)|^2 +
           x_2^2|\calM_2(\bar{r}_1,r_2)|^2 \right\}.
\label{QCDC1}
\eeqn
It is immediately clear that this is closely related to the \ee\ formula
(\ref{ee}), and is also an exact factorisation of the QCD and
electroweak parts of the process.

However there is one important difference from the \ee\ case: while the
parton-model configurations described by $r_1$ and $r_2$ are related by
a rotation in \ee, they are related by a transverse boost in the
scattering process.  This means that the simple Monte Carlo prescription
is not possible because although it seems natural to generate lowest
order \ee\ annihilation events with all possible orientations, it is not
so natural to generate scattering events with all possible intrinsic
transverse momenta.

It is worth noting that $x_2$ is not constrained to be positive, so the
vector $r_2$ can have a negative energy-component.  If one forgets this
fact, and simply evaluates the resulting matrix elements, then
(\ref{QCDC1}) works correctly, but to provide a physical interpretation,
one should first use $CPT$-invariance to rewrite the matrix element to
scatter a quark from $\bar{r}_1$ to $r_2$ as the matrix element to
scatter an antiquark from $-r_2$ to $-\bar{r}_1$.

In generating events, it is more convenient to introduce different
phase-space variables, $x_p\equiv-1/x_1$ and $z_p\equiv\frac{p_2\cdot
P}{q\cdot P}=1+(1-x_2)/x_1$, which have independent phase-space limits
\beqn
  x_\sssB \;<& x_p &<\; 1, \\
        0 \;<& z_p &<\; 1.
\eeqn
Inverting the relations we get
\beqn
  x_1 &=& -\frac{1}{x_p}, \\
  x_2 &=& 1-\frac{1-z_p}{x_p}.
\eeqn
We also make use of $x_\perp,$ the transverse momentum, measured in
units of $Q/2,$
\beq
  x_\perp^2 = \frac{4(1-x_p)(1-z_p)z_p}{x_p}
\eeq
to rewrite the expressions for $|C^\pm|^2$,
\beqn
  |C^+|^2 &=& \frac{8\pi\as}{(1-x_p)(1-z_p)Q^2} 
    \left\{x_p^2(x_2^2+x_\perp^2)\right\}
    \frac{x_2^2}{x_2^2+x_\perp^2}, \\
  |C^-|^2 &=& \frac{8\pi\as}{(1-x_p)(1-z_p)Q^2}
    \left\{1\right\}.
\eeqn
The reason for writing $|C^+|^2$ in this form is that as $x_2$ tends to
zero, the energy of $r_2$ tends to infinity, and the matrix element
$|\calM_2(\bar{r}_1,r_2)|^2$ in general diverges as $x_\perp^2/x_2^2,$
so the product
\beq
  \frac{x_2^2}{x_2^2+x_\perp^2}|\calM_2(\bar{r}_1,r_2)|^2
\eeq
remains finite.  The expressions in curly brackets in $|C^\pm|^2$ both
lie between zero and one throughout the physical phase-space.

The cross-section is then
\beqn
  d\sigma &=& \frac{C_\sssF\as}{2\pi}d\Gamma_2
    \frac{\myfrac{x_\sssB}{x_p}f(\myfrac{x_\sssB}{x_p},Q^2)}
         {        x_\sssB      f(        x_\sssB      ,Q^2)}
    \frac{dx_pdz_p}{(1-x_p)(1-z_p)}\;\frac{d\phi}{2\pi}
   \left\{ |\calM_2(r_1,\bar{r}_2)|^2 +
    \phantom{\frac{x_2^2}{x_2^2}}\right.\nonumber\\&&\left.
    \left(x_p^2(x_2^2+x_\perp^2)\right)
    \frac{x_2^2}{x_2^2+x_\perp^2}
    |\calM_2(\bar{r}_1,r_2)|^2 \right\}.
\label{QCDC2}
\eeqn
The product $d\Gamma_2|\calM_2(r_1,\bar{r}_2)|^2$ is exactly the lowest
order differential cross-section, so corresponding events are provided
by a parton-model event generator, but $|\calM_2(\bar{r}_1,r_2)|^2$
requires us to reevaluate the lowest order matrix element for the new
momenta.

The Monte Carlo procedure is then as follows:
\begin{enumerate}
\item
  Generate a parton model event according to the exact $\calO(1)$ matrix
  element.
\item
  Generate $x_p$ and $z_p$ values according to
  $dx_pdz_p/((1-x_p)(1-z_p))$.
\item
  Generate a $\phi$ value uniformly, and construct the corresponding
  momenta $p_1,\;p_2,\;k$ and $r_{1,2}$.
\item
  Calculate the ratio
\beq
  R_2 \equiv \frac{x_2^2}{x_2^2+x_\perp^2}
    \;\frac{|\calM_2(\bar{r}_1,r_2)|^2}{|\calM_2(r_1,\bar{r}_2)|^2}.
\eeq
\item
  Calculate a weight factor
\beq
  w\equiv\frac{C_\sssF\as}{2\pi}\;\frac
    {\myfrac{x_\sssB}{x_p}f(\myfrac{x_\sssB}{x_p},Q^2)}
    {        x_\sssB      f(        x_\sssB      ,Q^2)}\left\{
    1+x_p^2(x_2^2+x_\perp^2)R_2\right\}.
\eeq
\item
  Keep the event with probability proportional to $w$.
\end{enumerate}
This produces the {\em exact\/} distributions of $x_p,\;z_p$ and $\phi,$
for the given lowest-order phase-space point.  The average value of the
weight factor is the total $\calO(\as)$ correction to the cross-section
(at tree-level, according to the chosen cutoff).

This is as far as one can take the exact calculation for a general quark
scattering process, but if we now specialise to the DIS case, we can
proceed further.  The ratio $R_2$ can be quite generally written
\beq
  R_2 = \frac{\cos^2\theta_2 +
    \calA\cos\theta_2\left(l-\sqrt{l^2-1}\sin\theta_2\cos\phi\right) +
    \left(l-\sqrt{l^2-1}\sin\theta_2\cos\phi\right)^2}
    {1 + \calA l + l^2},
\eeq
where $\cos\theta_2=x_2/\sqrt{x_2^2+x_\perp^2}, \;
\sin\theta_2=x_\perp/\sqrt{x_2^2+x_\perp^2},$ and $l=2/y_\sssB-1,$ are
all Lorentz-invariant quantities that have simple interpretations in the
Breit frame: $\theta_2$ is the angle between $p_2$ and the exchanged
boson direction, and $Ql/2$ is the energy of the incoming lepton.
\calA\ is related to the couplings of the fermions to the exchanged
bosons,
\beq
  \calA = \frac{8C_{V,\ell}C_{A,\ell}C_{V,q}C_{A,q}}
               {(C_{V,\ell}^2+C_{A,\ell}^2)(C_{V,q}^2+C_{A,q}^2)}.
\eeq
For pure photon exchange $\calA_\gamma=0,$ for charged-current
interactions $\calA_{\sss{CC}}=2,$ and the full neutral-current case
including $\gamma$-Z interference can be calculated from
\beqn
  C_{V,i} &=& Q_i+(I_{3,i}/2-Q_i\sin^2\theta_w)
    \frac{Q^2}{(Q^2+m_{\sss Z}^2)\sin\theta_w\cos\theta_w}, \\
  C_{A,i} &=& \phantom{Q_i+(}I_{3,i}/2\phantom{{}-Q_i\sin^2\theta_w)}
    \frac{Q^2}{(Q^2+m_{\sss Z}^2)\sin\theta_w\cos\theta_w}.
\eeqn
Polarisation effects can also be incorporated by using appropriate
couplings in $\calA$.  Using these expressions we can take the azimuthal
average of $R_2$ analytically and generate the $\phi$ distribution
exactly, improving the Monte Carlo weight distribution.  It also allows
a direct interpretation of the azimuthal effects, as we discuss in a
later section.

\section{Boson Gluon Fusion}
\label{BGF}

The treatment of boson gluon fusion is mostly similar to QCD Compton, so
we do not go into as great detail.

The matrix element is
\beq
  \calM_3(\lambda,\rho) = ig_s\bar{u}_\lambda(p_2)
  \left[\slsh\epsilon_\rho(k)\frac{-(\slsh p_2-\slsh k)}{-2\ldot{p_2}{k}}
        \omega^\mu+\omega^\mu\frac{ (\slsh p_1-\slsh k)}{-2\ldot{p_1}{k}}
        \slsh\epsilon_\rho(k)\right]u_\lambda(p_1)J_\mu(q),
\eeq
where the momenta of the quark and antiquark are $p_2$ and $p_1$, and
all else is as before.  This can be simply obtained from the QCD Compton
case, by crossing
\beqn
  p_1 &\longrightarrow& -p_1, \\
  k   &\longrightarrow& -k.
\eeqn
The special vectors are
\beqn
\label{minus}
  r_1       &\equiv&          - p_1/x_1,\\
  \bar{r}_2 &\equiv& \phantom{-}r_1+q=p_2-k+p_1-p_1/x_1,\\
  r_2       &\equiv& \phantom{-}p_2/x_2,\\
  \bar{r}_1 &\equiv& \phantom{-}r_2-q=-p_1+k-p_2+p_2/x_2.
\eeqn
Note that $x_1$ now applies to the outgoing antiquark, rather than the
incoming quark.  This time, neither set corresponds to lowest-order
scattering---both have transverse components.  We obtain
\beqn
  |C^+|^2 &=& \frac{8\pi\as}{(1-x_1)(1-x_2)Q^2} x_2^2 ,\\
  |C^-|^2 &=& \frac{8\pi\as}{(1-x_1)(1-x_2)Q^2} x_1^2 .
\eeqn

We again parametrise the kinematics by $x_p$ and $z_p,$ where $z_p$
still refers to the outgoing quark, ie.~particle 2, and $x_p$ refers to
the incoming particle, which is this time the gluon.  We then obtain
\beqn
  x_1 &=& 1-\frac{z_p}{x_p}, \\
  x_2 &=& 1-\frac{1-z_p}{x_p}.
\eeqn
This gives
\beqn
  |C^+|^2 &=& \frac{8\pi\as}{z_p(1-z_p)Q^2} 
    \left\{x_p^2(x_2^2+x_\perp^2)\right\}
    \frac{x_2^2}{x_2^2+x_\perp^2}, \\
  |C^-|^2 &=& \frac{8\pi\as}{z_p(1-z_p)Q^2}
    \left\{x_p^2(x_1^2+x_\perp^2)\right\}
    \frac{x_1^2}{x_1^2+x_\perp^2}.
\eeqn
The cross-section is then
\beqn
  d\sigma
  &=& \frac{\half\as}{2\pi}d\Gamma_2
    \frac{\myfrac{x_\sssB}{x_p}f_g(\myfrac{x_\sssB}{x_p},Q^2)}
         {        x_\sssB      f_q(        x_\sssB      ,Q^2)}
    \frac{dx_pdz_p}{z_p(1-z_p)}\;\frac{d\phi}{2\pi}
\nonumber\\&&
  \left\{
    \left(x_p^2(x_1^2+x_\perp^2)\right)R_1+
    \left(x_p^2(x_2^2+x_\perp^2)\right)R_2\right\}
    |\calM_2(q_1,q_2)|^2,
\eeqn
where $q_{1,2}$ are the parton-model momenta, and
\beqn
  R_1 &=& \frac{x_1^2}{x_1^2+x_\perp^2}
    \;\frac{|\calM_2(r_1,\bar{r}_2)|^2}{|\calM_2(q_1,q_2)|^2}, \\
  R_2 &=& \frac{x_2^2}{x_2^2+x_\perp^2}
    \;\frac{|\calM_2(\bar{r}_1,r_2)|^2}{|\calM_2(q_1,q_2)|^2},
\eeqn
which can be rewritten for the explicit case of DIS as before.  We
obtain the identical expression for $R_2,$ and the same for $R_1,$ but
with $\theta_2$ and $\phi$ replaced by the corresponding $\theta_1$
and $\pi-\phi$.

Up to here, the BGF cross-section has been written purely in terms of
the lowest-order quark scattering (ie.~not antiquarks).  There is
nothing wrong with this in itself, as the exact distributions of BGF
events will be correctly reproduced.  However, if we want to view this
as a correction to the lowest-order process, rather than just a
cross-section calculation, it would certainly be more desirable to treat
quarks and antiquarks equivalently.  We can do this by noting that the
$z_p=1$ singularity is associated with configurations that become
collinear to lowest-order quark scattering, while $z_p=0$ is associated
with antiquark scattering.  We use the separation
\beq
  \frac1{z_p(1-z_p)} = \frac1{z_p} + \frac1{1-z_p}
\eeq
to rewrite the cross-section
\beqn
  d\sigma
  &=& \frac{\half\as}{2\pi}d\Gamma_2
    \frac{\myfrac{x_\sssB}{x_p}f_g(\myfrac{x_\sssB}{x_p},Q^2)}
         {        x_\sssB      f_q(        x_\sssB      ,Q^2)}
    \frac{dx_pdz_p}{1-z_p}\;\frac{d\phi}{2\pi}
\nonumber\\&&\left\{
    \left(x_p^2(x_1^2+x_\perp^2)\right)R_1+
    \left(x_p^2(x_2^2+x_\perp^2)\right)R_2
    \right\}|\calM_2(q_1,q_2)|^2
\nonumber\\
  &+& \frac{\half\as}{2\pi}d\Gamma_2
    \frac{\myfrac{x_\sssB}{x_p}f_g(\myfrac{x_\sssB}{x_p},Q^2)}
         {        x_\sssB      f_q(        x_\sssB      ,Q^2)}
    \frac{dx_pdz_p}{z_p}\;\frac{d\phi}{2\pi}
\nonumber\\&&\left\{
    \left(x_p^2(x_1^2+x_\perp^2)\right)R_1+
    \left(x_p^2(x_2^2+x_\perp^2)\right)R_2
    \right\}|\calM_2(q_1,q_2)|^2.
\eeqn
Now we define a set of exchanged variables,
\beqn
  \widetilde{p}_1 &\equiv& p_2, \\
  \widetilde{p}_2 &\equiv& p_1, \\
  \widetilde{z}_p &\equiv& 1-z_p.
\eeqn
Although the special vectors are defined in the same way, they become
negated, because of the minus sign in (\ref{minus}), so we obtain
\beqn
  d\sigma
  &=& \frac{\half\as}{2\pi}d\Gamma_2
    \frac{\myfrac{x_\sssB}{x_p}f_g(\myfrac{x_\sssB}{x_p},Q^2)}
         {        x_\sssB      f_q(        x_\sssB      ,Q^2)}
    \frac{dx_pdz_p}{1-z_p}\;\frac{d\phi}{2\pi}
\nonumber\\&&\left\{
    \left(x_p^2(x_1^2+x_\perp^2)\right)R_1+
    \left(x_p^2(x_2^2+x_\perp^2)\right)R_2
    \right\}|\calM_2(q_1,q_2)|^2
\nonumber\\
  &+& \frac{\half\as}{2\pi}d\Gamma_2
    \frac{\myfrac{x_\sssB}{x_p}f_g(\myfrac{x_\sssB}{x_p},Q^2)}
         {        x_\sssB      f_q(        x_\sssB      ,Q^2)}
    \frac{dx_pd\widetilde{z}_p}{1-\widetilde{z}_p} \;
      \frac{d\widetilde\phi}{2\pi}
\nonumber\\&&\left\{
    \left(x_p^2(\widetilde{x}_2^2+x_\perp^2)\right)
    \frac{\widetilde{x}_2^2}{\widetilde{x}_2^2+x_\perp^2}
    |\calM_2(-\widetilde{r}_2,-\widetilde{\bar{r}}_1)|^2
\right.\nonumber\\&&\left.+
    \left(x_p^2(\widetilde{x}_1^2+x_\perp^2)\right)
    \frac{\widetilde{x}_1^2}{\widetilde{x}_1^2+x_\perp^2}
    |\calM_2(-\widetilde{\bar{r}}_2,-\widetilde{r}_1)|^2
    \right\}.
\eeqn
Finally, we can use the $CPT$-invariance of the matrix element, to write
\beq
  |\calM_2(-q_2,-q_1)|^2 = |\widetilde\calM_2(q_1,q_2)|^2,
\eeq
where $\widetilde\calM_2(q_1,q_2)$ is the matrix element to scatter an
antiquark from $q_1$ to $q_2$.  Thus the two halves of the cross-section
are identical, but with quarks and antiquarks interchanged,
\beqn
  d\sigma
  &=& \frac{\half\as}{2\pi}d\Gamma_2
    \frac{\myfrac{x_\sssB}{x_p}f_g(\myfrac{x_\sssB}{x_p},Q^2)}
         {        x_\sssB      f_q(        x_\sssB      ,Q^2)}
    \frac{dx_pdz_p}{1-z_p}\;\frac{d\phi}{2\pi}
\nonumber\\&&\left\{
    \left(x_p^2(x_1^2+x_\perp^2)\right)R_1+
    \left(x_p^2(x_2^2+x_\perp^2)\right)R_2
    \right\}|\calM_2(q_1,q_2)|^2
\nonumber\\
  &+& \frac{\half\as}{2\pi}d\Gamma_2
    \frac{\myfrac{x_\sssB}{x_p}f_g(\myfrac{x_\sssB}{x_p},Q^2)}
         {        x_\sssB      f_{\bar q}(        x_\sssB      ,Q^2)}
    \frac{dx_pd\widetilde{z}_p}{1-\widetilde{z}_p}\;\frac{d\widetilde\phi}{2\pi}
\nonumber\\&&\left\{
    \left(x_p^2(\widetilde{x}_1^2+x_\perp^2)\right)\widetilde{R}_1+
    \left(x_p^2(\widetilde{x}_2^2+x_\perp^2)\right)\widetilde{R}_2
    \right\}|\widetilde\calM_2(q_1,q_2)|^2.
\label{BGF1}
\eeqn
The two halves can then be separately associated with lowest-order
scattering of quarks and antiquarks.

The Monte Carlo algorithm is then:
\begin{enumerate}
\item
  Generate a parton model quark or antiquark event according to the
  exact $\calO(1)$ matrix element.
\item
  Generate $x_p$ and $z_p$ values according to $dx_pdz_p/(1-z_p)$.
\item
  Generate a $\phi$ value uniformly, and construct the corresponding
  momenta $p_1,\;p_2,\;k$ and $r_{1,2}$.
\item
  Calculate the ratios $R_1$ and $R_2$.
\item
  Calculate a weight factor
\beq
  w = \frac{\half\as}{2\pi}\frac
    {\myfrac{x_\sssB}{x_p}f_g         (\myfrac{x_\sssB}{x_p},Q^2)}
    {        x_\sssB      f_{q/\bar q}(        x_\sssB      ,Q^2)}
  \left\{x_p^2(x_1^2+x_\perp^2)R_1+x_p^2(x_2^2+x_\perp^2)R_2\right\}.
\eeq
\item
  Keep the event with probability proportional to $w$.
\end{enumerate}
This again produces the {\em exact\/} distributions at the given
phase-space point, and gives the total tree-level $\calO(\as)$
correction there.

\section{Lepton-Hadron Correlations in DIS}
\label{azi}

Before discussing the azimuthal correlations that arise in first order
QCD, it is worth recalling that a correlation also arises in the parton
model, if the parton's intrinsic $p_t$ in a hadron is included.  This is
because at fixed $Q^2,$ different azimuths correspond to different
parton-lepton collision energies, so even though the Fermi motion of
partons in a hadron is azimuthally uniform, their scattering
cross-section is not.  The matrix element is identical to
$|\calM_2(\bar{r}_1,r_2)|^2$ evaluated earlier,
\beq
  |\calM|^2
  \propto
  \frac{\cos^2\theta_2 +
    \calA\cos\theta_2\left(l-\sqrt{l^2-1}\sin\theta_2\cos\phi\right) +
    \left(l-\sqrt{l^2-1}\sin\theta_2\cos\phi\right)^2}
    {1 + \calA l + l^2}.
\eeq
It is this partonic matrix element that determines the azimuthal
asymmetry[\ref{C}], and not the corresponding partonic cross-section
($\sim|\calM|^2/s,$ where $s$ is the parton-lepton invariant mass).
$\theta_2$ can be directly calculated since in the Breit frame, the
incoming parton has longitudinal momentum $Q/2$ and transverse momentum
$p_t,$ so we have
\beq
  \tan\theta_2 = 2p_t/Q.
\eeq
The size of the correlation is usually parametrised by the average
values of $\cos\phi$ and $\cos2\phi,$
\beqn
  \VEV{\cos\phi} &=&
    \frac{-\calA\cos\theta_2\sqrt{l^2-1}\sin\theta_2 -
    2l\sqrt{l^2-1}\sin\theta_2}
    {2(\cos^2\theta_2+\calA\cos\theta_2l+l^2+\myfrac12(l^2-1)\sin^2\theta_2)},
  \\
  \VEV{\cos2\phi} &=&
    \frac{\myfrac12(l^2-1)\sin^2\theta_2}
    {2(\cos^2\theta_2+\calA\cos\theta_2l+l^2+\myfrac12(l^2-1)\sin^2\theta_2)}.
\eeqn
It can be seen that both effects are maximised by working at small
$Q^2,$ so that $\theta_2$ is maximised (though it is still small for
accessible values of $Q^2$).  It is also increased somewhat by working
at small $y_\sssB$ (ie.~$l\gg1$).  Taking the limit $p_t\ll Q$ and
using the definition of $l,$ we obtain the usual expressions[\ref{C}],
\beqn
  \VEV{\cos\phi} &=&
    \frac{(2-y_\sssB + \myfrac12\calA y_\sssB)\sqrt{1-y_\sssB}}
    {1+(1-y_\sssB)^2 + \myfrac12\calA y_\sssB(2-y_\sssB)}
    \left(\frac{-2p_t}{Q}\right),
  \\
  \VEV{\cos2\phi} &=&
    \frac{2(1-y_\sssB)}
    {1+(1-y_\sssB)^2 + \myfrac12\calA y_\sssB(2-y_\sssB)}
    \left(\frac{p_t^2}{Q^2}\right).
\eeqn
In the small-$y_\sssB$ limit, these cross-sections become independent of
$\calA$ and, as noted in [\ref{C}] are even the same if the fermions are
replaced by scalars.  This corresponds to the well-known fact that the
cross-section for scattering by exchange of a single spin-1 particle has
the same high-energy behaviour,
\beq
  |\calM|^2
  \propto
  \left(\frac{s}{Q^2}\right)^2,
\eeq
independent of its vector/axial-vector nature, its couplings, and the
particles being scattered.  This is the dominant contribution, even when
$y_\sssB$ is not small.

Turning now to QCDC, we recall that the total cross-section is the
weighted sum of parton-model pieces where there is no $p_t,$ and where
there is a $p_t$ of
\beq
  p_t^2 = \myfrac14Q^2\frac{x_\perp^2}{x_2^2}
    = Q^2\frac{(1-x_p)(1-z_p)x_pz_p}{(x_p+z_p-1)^2},
\label{pt2}
\eeq
whose relative importance is reduced by $Q^2/(Q^2+4p_t^2)$.

Studying the first-order cross-section (\ref{QCDC1}), it is clear that
only $|\calM_2(\bar{r}_1,r_2)|^2$ generates any $\phi$-dependence, and
that this is identical to that in the parton model with intrinsic $p_t$
given by (\ref{pt2}).  One can therefore say that QCD does not play a
{\em direct\/} part in determining the azimuthal correlations.  Its
r\^ole can be summarised as providing a large `intrinsic' $p_t,$ and
then diluting the resulting correlation.  Furthermore, as mentioned
above, this is dominated by the fact that the exchanged particle has
spin-1.  This also applies to the BGF part, although the dilution is not
so great, because both contributions have an `intrinsic' $p_t$.

We can calculate the size of the correlation,
\beqn
  \VEV{\cos\phi} &=&
    \frac{\int d\sigma_3\cos\phi}{\int d\sigma_3},
  \\
  \VEV{\cos2\phi} &=&
    \frac{\int d\sigma_3\cos2\phi}{\int d\sigma_3},
\eeqn
where the integrals are over whatever phase-space region is selected for
the analysis.  We examine two regions in particular, firstly the point
where the correlation is maximised.  Studying (\ref{QCDC2}), we find
that both \VEV{\cos\phi} and \VEV{\cos2\phi} are maximised at
$x_p=z_p=1/2$.  This corresponds to scattering in which both outgoing
partons are at $90^\circ$ to the boson-hadron axis in the Breit frame,
with energy $Q/2,$ so $\surd\hat{s}=2p_t=Q$.  Going to higher $p_t$
increases the size of the correlation generated by
$|\calM_2(\bar{r}_1,r_2)|^2$, but also increases the dilution, so that
the correlation gets smaller overall.  At that point we have
\beqn
  \VEV{\cos\phi}_{\sss{QCDC,max}} &=&
    \frac{-2\sqrt{1-y_\sssB}(2-y_\sssB)}
    {9(1+(1-y_\sssB)^2)+4(1-y_\sssB)+4\calA y_\sssB(2-y_\sssB)},
  \\
  \VEV{\cos2\phi}_{\sss{QCDC,max}} &=&
    \frac{\myfrac14(2-y_\sssB)^2}
    {9(1+(1-y_\sssB)^2)+4(1-y_\sssB)+4\calA y_\sssB(2-y_\sssB)}.
\eeqn
The BGF case is not quite so simple, because although the contributions
to \VEV{\cos\phi} from the $R_1$ and $R_2$ terms in (\ref{BGF1}) are
separately maximised at $x_p=z_p=1/2,$ they are equal and opposite
there, so exactly cancel.  The sum is maximised at $x_p=z_p =
(2+\sqrt{8\surd3-12})/4 \approx 0.841,$ which corresponds to a
configuration where one of the partons is at $90^\circ$ to the
boson-hadron axis, and the other is about $20^\circ$ from the current
direction, with $p_t\approx0.159Q$ and $\surd\hat s\approx0.435Q$.  The
value at that point does not have a simple form, but is numerically
similar to $\VEV{\cos\phi}_{\sss{QCDC,max}}$.  It would not be as easy
to measure though, since different parts of phase-space give
correlations with different signs, so integrating over some region of
phase-space reduces \VEV{\cos\phi} faster for BGF than for QCDC.  For
\VEV{\cos2\phi}, the maximum is again at $x_p=z_p=1/2,$ with value
\beq
  \VEV{\cos2\phi}_{\sss{BGF,max}} =
    \frac{1-y_\sssB}{1+(1-y_\sssB)^2+4(1-y_\sssB)},
\eeq
independent of \calA, almost four times larger than
$\VEV{\cos2\phi}_{\sss{QCDC,max}},$ and comparable to
$\VEV{\cos\phi}_{\sss{QCDC,max}}$.

As already mentioned, as the transverse momentum is increased above
$Q/2,$ the correlation becomes weaker again.  This is because the
correlation generated by $|\calM_2(\bar{r}_1,r_2)|^2$ is governed by the
direction of $p_2$ in the Breit frame, rather than its $p_t$.  Thus the
correlation cannot get any stronger than it is when $p_2$ is at
$90^\circ$ to the current direction.  However, the dilution by the
azimuthally flat term increases with increasing $\hat{s},$ so the
correlation decreases overall.  Since most experimental analyses use a
fixed cutoff in $p_t$ irrespective of $Q,$ it is primarily this
dependence that determines the strong overall dependence on $Q^2$.

The other phase-space we consider is to not make any cuts at all.  This
is possible because both \VEV{\cos\phi} and \VEV{\cos2\phi} are infrared
safe quantities, so if $\phi$ is measured semi-inclusively for all DIS
events, the expectation values can be expanded as simple power series in
\as\footnote{This is only strictly true if we define $\phi$ in such a
way that the expectation values are identical in the $p_t\to0$ limit and
at $p_t=0,$ ie.~if they are zero in the latter case.  This can be
achieved by adding the rule that if no particles in the event have any
momentum transverse to the boson-hadron axis, a $\phi$ value is chosen
randomly and uniformly.  Since experimentally such events never exist,
this makes no practical difference to the result.}.  We can then write
the expectations symbolically as
\beq
  \VEV{\cos\phi} = \frac{\int d\sigma_3\cos\phi}
    {1+\int d\sigma_3-C\as},
\eeq
with an equivalent expression for \VEV{\cos2\phi}.  Because the
integrals are now over the whole phase-space, the integral in the
denominator is divergent, but $C,$ the first-order virtual correction to
the lowest-order result, is also divergent such that after
renormalisation and factorisation their sum is the finite first-order
QCD correction to the total DIS cross-section.  The integral in the
numerator is convergent.  Since it is first-order in \as, we can ignore
the \as\ dependence of the denominator to first order, and we obtain
simply
\beq
  \VEV{\cos\phi} = \int d\sigma_3\cos\phi.
\eeq
\pagebreak[3]\par
\begin{figure}[tb]
  \vspace*{6.5cm}
  \includegraphics{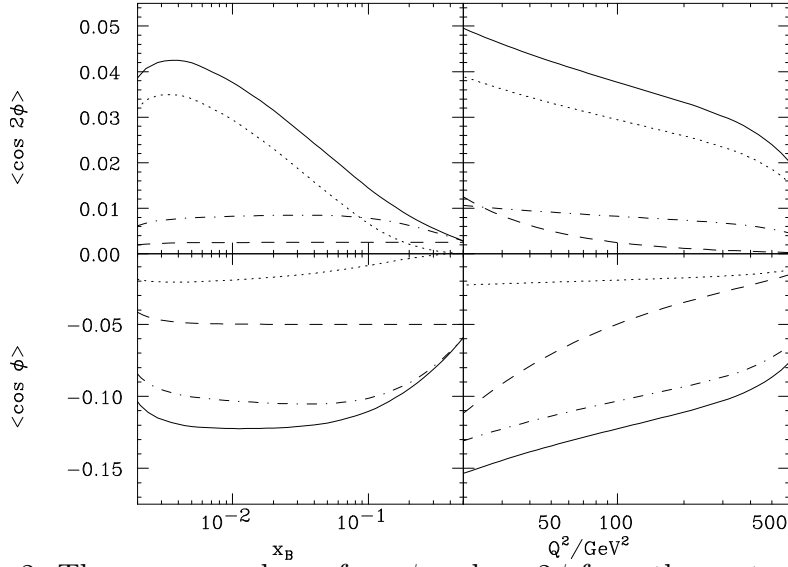}
  \caption[]{The average values of $\cos\phi$ and $\cos2\phi$ from the
    parton model (dashed) and QCD (solid), for ep collisions at
    $s=10^5\,\mathrm{GeV^2}$ with $Q^2=10^2\,\mathrm{GeV^2}$ (left)
    and $x_\sssB=0.01$ (right).  We use the MRS D$-'$ distribution
    functions, $\Lambda_{\sss{QCD}}=230\,$MeV, with pure photon
    exchange, and $\VEV{p_t}=\surd\!\VEV{p_t^2}=500\,$MeV.  The QCD
    curves are also broken down into the separate contributions from
    QCDC (dot-dashed) and BGF (dotted).}
\label{correl}
\end{figure}

\noindent
For QCDC we obtain
\beqn
  \VEV{\cos\phi}_{\sss{QCDC}} &=& -\frac{C_\sssF\as}{4\pi}
    \int_{x_\sssB}^1dx_p \int_0^1dz_p \frac1{(1-x_p)(1-z_p)} \frac
    {\myfrac{x_\sssB}{x_p}f(\myfrac{x_\sssB}{x_p},Q^2)}
    {        x_\sssB      f(        x_\sssB      ,Q^2)}
\nonumber\\&&
    x_p^2(x_2^2+x_\perp^2)
    \frac{\calA\sqrt{l^2-1}\cos\theta_2\sin\theta_2 +
      2l\sqrt{l^2-1}\sin\theta_2}{1+\calA l+l^2},
\eeqn
which can be integrated over $z_p$ to give
\beqn
  \VEV{\cos\phi}_{\sss{QCDC}} &=& -\frac{C_\sssF\as}{16}
    \int_{x_\sssB}^1dx_p \sqrt{\frac{x_p}{1-x_p}} \frac
    {\myfrac{x_\sssB}{x_p}f(\myfrac{x_\sssB}{x_p},Q^2)}
    {        x_\sssB      f(        x_\sssB      ,Q^2)}
\nonumber\\&&
    \frac{\calA\sqrt{l^2-1}(4x_p-1) +
      2l\sqrt{l^2-1}(1+2x_p)}{1+\calA l+l^2},
\eeqn
which must be integrated numerically.  We similarly obtain
\beq
  \VEV{\cos2\phi}_{\sss{QCDC}} = \frac{C_\sssF\as}{4\pi}
    \frac{l^2-1}{1+\calA l+l^2}
    \int_{x_\sssB}^1dx_p x_p \frac
    {\myfrac{x_\sssB}{x_p}f(\myfrac{x_\sssB}{x_p},Q^2)}
    {        x_\sssB      f(        x_\sssB      ,Q^2)},
\eeq
and
\beqn
  \VEV{\cos\phi}_{\sss{BGF}} &=& -\frac{\myfrac12\as}{8}
    \int_{x_\sssB}^1dx_p \sqrt{x_p(1-x_p)} \frac
    {\myfrac{x_\sssB}{x_p}f_g(\myfrac{x_\sssB}{x_p},Q^2)}
    {        x_\sssB      f_q(        x_\sssB      ,Q^2)}
\nonumber\\&&
    \frac{\calA\sqrt{l^2-1} +
      2l\sqrt{l^2-1}(2x_p-1)}{1+\calA l+l^2}, \\
  \VEV{\cos2\phi}_{\sss{BGF}} &=& \frac{\myfrac12\as}{2\pi}
    \frac{l^2-1}{1+\calA l+l^2}
    \int_{x_\sssB}^1dx_p x_p (1-x_p)\frac
    {\myfrac{x_\sssB}{x_p}f_g(\myfrac{x_\sssB}{x_p},Q^2)}
    {        x_\sssB      f_q(        x_\sssB      ,Q^2)}.
\eeqn
Note that these are the extreme values of the expectations, assuming
that the scattered parton direction could be perfectly identified.  They
would be reduced by a realistic method, such as using the hadron with
largest Feynman-$x_\sssF,$ or using all particles weighted by $x_\sssF$.
The values of these expectations are shown in Fig.~\ref{correl}, in
comparison with those from the intrinsic $p_t$ in the parton model.

Although we have seen that the physical origin of the correlation is the
same in QCD as in the parton model, the dependence on $x_\sssB$ and
$Q^2$ is quite difference, so the two contributions can be easily
separated.

\section{The Drell-Yan Process}
\label{DY}

The treatment of the Drell-Yan process again proceeds along similar
lines.  We start with the annihilation+gluon process,
$\mathrm{q\bar{q}\to gV^*}$, for which the matrix element is
\beq
  \calM_3(\lambda,\rho) = ig_s\bar{v}_\lambda(p_2)
  \left[\slsh\epsilon_\rho(k)\frac{ (\slsh p_2-\slsh k)}{-2\ldot{p_2}{k}}
        \omega^\mu+\omega^\mu\frac{-(\slsh p_1-\slsh k)}{-2\ldot{p_1}{k}}
        \slsh\epsilon_\rho(k)\right]u_\lambda(p_1)J_\mu(q),
\eeq
where the momenta of the quark and antiquark are $p_1$ and $p_2$.  This
can again be obtained from the QCD Compton case, by crossing
\beq
  p_2 \longrightarrow -p_2.
\eeq
The special vectors are
\beqn
  r_1       &\equiv& p_1/x_1,\\
  \bar{r}_2 &\equiv& q-r_1=p_2-k+p_1-p_1/x_1,\\
  r_2       &\equiv& p_2/x_2,\\
  \bar{r}_1 &\equiv& q-r_2=p_1-k+p_2-p_2/x_2.
\eeqn
We then obtain
\beqn
  |C^+|^2 &=& \frac{8\pi\as}{(x_1-1)(x_2-1)Q^2} x_2^2 ,\\
  |C^-|^2 &=& \frac{8\pi\as}{(x_1-1)(x_2-1)Q^2} x_1^2 .
\eeqn
The cross-section in lowest order is
\beq
  d\sigma_2 = \frac1sdQ^2dy\; f_{q/1}(\eta_1)f_{\bar{q}/2}(\eta_2)
              \frac1{2Q^2} |\calM_2|^2d\Gamma_2,
\eeq
where $y$ is the rapidity of the gauge boson, $\eta_{1,2}=Q/\surd
se^{\pm y}$ are the energy-fractions of the quarks in the hadrons,
$f_{q/h}$ are the corresponding distribution functions, and $d\Gamma_2$
is an element of the phase-space for the gauge boson decay.  The
corresponding first-order cross-section is
\beqn
  d\sigma_3 &=& \frac1sd\hat{s}dy\; f_{q/1}(\xi_1)f_{\bar{q}/2}(\xi_2)
                \frac1{2Q^2}\;\frac{C_\sssF\as}{2\pi}\;
                \frac{dx_1dx_2}{(x_1+x_2-1)^3(x_1-1)(x_2-1)}\;
                \frac{d\phi}{2\pi}
\nonumber\\&&
   \left\{ x_1^2|\calM_2(r_1,\bar{r}_2)|^2 +
           x_2^2|\calM_2(\bar{r}_1,r_2)|^2 \right\}
              d\Gamma_2.
\eeqn
Just as in the \ee\ case, this can be evaluated exactly, without any
knowledge of $|\calM_2|^2$.  The event can be boosted from the lab frame
to the rest frame of the boson in such a way that the azimuth of the
gluon around either $p_1$ or $p_2$ is identical to its lab-frame azimuth
around the beam direction.  However, these two different possibilities
correspond to different configurations of the boson decay in the lab
frame.  $|\calM_2(r_1,\bar{r}_2)|^2$ clearly has no dependence on the
azimuth around $p_1,$ and $|\calM_2(\bar{r}_1,r_2)|^2$ has no dependence
on the azimuth around $p_2$.  This means that the boson decays can be
correctly generated by generating a uniform azimuth around one or other
of the quarks in the boson's rest frame, just as in \ee\ annihilation.

It then only remains to relate the two cross-sections.  Switching to
using $Q^2$ and $\hat{t}=-2\ldot{p_1}{k}=-Q^2(x_2-1)$ to parametrise
the hadron-plane momenta, we obtain
\beqn
  d\sigma_3 &=& \frac1sd\hat{s}dydQ^2d\hat{t}\;
                f_{q/1}(\xi_1)f_{\bar{q}/2}(\xi_2)\;
                \frac1{2Q^2}\;\frac{C_\sssF\as}{2\pi}\;
                \frac1{\hat{s}^2\hat{t}\hat{u}}\;
                \frac{d\phi}{2\pi}
\nonumber\\&&
   \left\{ (Q^2-\hat{u})^2|\calM_2(r_1,\bar{r}_2)|^2 +
           (Q^2-\hat{t})^2|\calM_2(\bar{r}_1,r_2)|^2 \right\}
              d\Gamma_2
\eeqn
and hence
\beq
  d\sigma_3 =   \frac{f_{q/1}(\xi_1 )f_{\bar{q}/2}(\xi_2 )}
                     {f_{q/1}(\eta_1)f_{\bar{q}/2}(\eta_2)}
                \frac{C_\sssF\as}{2\pi}\;
                \frac{d\hat{s}d\hat{t}}
                     {\hat{s}^2\hat{t}\hat{u}}
   \left\{ (Q^2-\hat{u})^2 d\sigma_2\frac{d\phi}{2\pi} + 
           (Q^2-\hat{t})^2 d\sigma_2\frac{d\phi}{2\pi} \right\}.
\eeq
%
The Monte Carlo algorithm is then
\begin{enumerate}
\item
  Generate a parton model event according to the exact $\calO(1)$ matrix
  element.
\item
  Generate $\hat{s}$ and $\hat{t}$ values according to
  $Q^4d\hat{s}d\hat{t}/\hat{s}^2\hat{t}\hat{u} =
  Q^4d\hat{s}d\hat{t}/\hat{s}^2\hat{t}(Q^2-\hat{s}-\hat{t})$.
\item
  Construct momenta in the boson rest-frame corresponding to these values.
\item
  Choose parton 1 with probability $(Q^2-\hat{u})^2 /
  \left((Q^2-\hat{u})^2+(Q^2-\hat{t})^2\right),$ and otherwise parton 2,
  and uniformly rotate about the chosen parton.
\item
  Boost the momenta back to the lab frame such that the boson's rapidity
  is the same as it was in the parton-model event.
\item
  Calculate a weight factor
\beq
  w = \frac{C_\sssF\as}{2\pi}\frac
      {f_{q/1}(\xi_1 )f_{\bar{q}/2}(\xi_2 )}
      {f_{q/1}(\eta_1)f_{\bar{q}/2}(\eta_2)}
  \left\{(1-\hat{u}/Q^2)^2+(1-\hat{t}/Q^2)^2\right\}.
\eeq
\item
  Keep the event with probability proportional to $w$.
\end{enumerate}
This correctly gives the properties of annihilation+gluon events, for
example correlations between the gluon and lepton directions, with no
knowledge of the matrix element that determines them, provided a
generator of lowest-order events is available.

Turning to the Compton process, $\mathrm{qg\to qV^*}$, we find that
exactly the same treatment holds, and we obtain
\beq
  d\sigma_3 =   \frac{f_{q/1}(\xi_1 )f_{     g /2}(\xi_2 )}
                     {f_{q/1}(\eta_1)f_{\bar{q}/2}(\eta_2)}
                \frac{\myfrac12\as}{2\pi}\;
                \frac{d\hat{s}d\hat{t}}
                     {-\hat{s}^3\hat{t}}
   \left\{ (Q^2-\hat{t})^2 d\sigma_2\frac{d\phi}{2\pi} + 
           (Q^2-\hat{s})^2 d\sigma_2\frac{d\phi}{2\pi} \right\},
\eeq
where $\hat{t}$ is this time $-2\ldot{p_2}{k}$.  Equivalent formul\ae\
hold when the quark is replaced by an antiquark, or comes from the other
hadron.

The Monte Carlo algorithm is then
\begin{enumerate}
\item
  Generate a parton model event according to the exact $\calO(1)$ matrix
  element.
\item
  Generate $\hat{s}$ and $\hat{t}$ values according to
  $-Q^4d\hat{s}d\hat{t}/\hat{s}^3\hat{t}$.
\item
  Construct momenta in the boson rest-frame corresponding to these values.
\item
  Choose parton 1 with probability $(Q^2-\hat{t})^2 /
  \left((Q^2-\hat{t})^2+(Q^2-\hat{s})^2\right),$ and otherwise parton 2,
  and uniformly rotate about the chosen parton.
\item
  Boost the momenta back to the lab frame such that the boson's rapidity
  is the same as it was in the parton-model event.
\item
  Calculate a weight factor
\beq
  w = \frac{\myfrac12\as}{2\pi}\frac
      {f_{q/1}(\xi_1 )f_{     g /2}(\xi_2 )}
      {f_{q/1}(\eta_1)f_{\bar{q}/2}(\eta_2)}
  \left\{(1-\hat{t}/Q^2)^2+(1-\hat{s}/Q^2)^2\right\}.
\eeq
\item
  Keep the event with probability proportional to $w$.
\end{enumerate}
This again gives the correct event properties, with no knowledge of the
matrix element that determines them, provided a generator of
lowest-order events is available.

\section{Summary}
\label{summ}

By working in the simple gauge introduced by the CALKUL collaboration,
it is possible to {\em exactly} factorise the first order corrections to
the electroweak production, scattering and annihilation of quarks.  This
allows a transparent understanding of event properties and correlations,
and enables simple Monte Carlo prescriptions to be constructed.  This
was done in~[\ref{K}] for production processes, and in the present paper
for scattering and annihilation processes.

For production and annihilation the factorisation is complete, in the
sense that the Monte Carlo algorithm does not need any details of the
lowest-order cross-section to generate the correction.  This is because
it is able to use the angular configurations generated at lowest order,
provided that all such configurations are generated.  In the scattering
case, the equivalent requirement is that all transverse boosts are
generated, which is not a natural situation.  Instead, the lowest order
matrix element must be reevaluated for a new set of momenta.

As an example of the additional insight that our approach can give, we
discussed azimuthal correlations between the leptonic and hadronic
planes in DIS.  It was seen that the correlation that arises in
first-order QCD has the same physical origin as that in the parton
model---simply the variation of the parton-lepton invariant mass with
azimuth, when there is a non-zero transverse momentum.  The r\^ole of
QCD is simply to provide that transverse momentum.  By separating the
cross-section into two components, our method is able to describe this
transverse momentum in a gauge-invariant way.

We finally note that since our results for first-order cross-sections
are explicitly formulated as corrections to lowest-order cross-sections,
they are ideally suited for making first-order matrix-element
corrections to parton shower algorithms[\ref{S}].

\section*{References}
\begin{enumerate}
\item\label{K}
  R.~Kleiss, Phys.~Lett.~180B (1986) 400.
\item\label{GP}
  H.~Georgi, H.D.~Politzer, Phys.~Rev.~Lett.~40 (1978) 3.
\item\label{CES}
  J.~Chay, S.D.~Ellis, W.J.~Stirling, Phys.~Rev.~D45 (1992) 46.
\item\label{CALKUL}
  The CALKUL Collaboration, Nucl.~Phys.~B206 (1982) 53.
\item\label{C}
  R.N.~Cahn, Phys.~Lett.~78B (1978) 269. \\
  R.N.~Cahn, Phys.~Rev.~D40 (1989) 3107.
\item\label{S}
  M.H.~Seymour, LU TP 94-12, contributed to the 27th International
  Conference on High Energy Physics, Glasgow, U.K., 20--27 July 1994. \\
  M.H.~Seymour, {\em Matrix-Element Corrections to Angular-Ordered
  Parton Showers,} in preparation.
\end{enumerate}

\end{document}